# Structural Diversity and Homophily: A Study Across More Than One Hundred Big Networks

Yuxiao Dong*
Microsoft Research
Redmond, WA 98052
yuxdong@microsoft.com

Reid A. Johnson
University of Notre Dame
Notre Dame, IN 46556
rjohns15@nd.edu

Jian Xu
University of Notre Dame
Notre Dame, IN 46556
jxu5@nd.edu

Nitesh V. Chawla
University of Notre Dame
Notre Dame, IN 46556
nchawla@nd.edu

## ABSTRACT

A widely recognized organizing principle of networks is structural homophily, which suggests that people with more common neighbors are more likely to connect with each other. However, what influence the diverse structures embedded in common neighbors (e.g.,  and ) have on link formation is much less well-understood. To explore this problem, we begin by characterizing the structural diversity of common neighborhoods. Using a collection of 120 large-scale networks, we demonstrate that the impact of the common neighborhood diversity on link existence can vary substantially across networks. We find that its positive effect on Facebook and negative effect on LinkedIn suggest different underlying networking needs in these networks. We also discover striking cases where diversity violates the principle of homophily—that is, where fewer mutual connections may lead to a higher tendency to link with each other. We then leverage structural diversity to develop a common neighborhood signature (CNS), which we apply to a large set of networks to uncover unique network superfamilies not discoverable by conventional methods. Our findings shed light on the pursuit to understand the ways in which network structures are organized and formed, pointing to potential advancement in designing graph generation models and recommender systems.

## CCS CONCEPTS

•**Information systems** →**Social networks;** •**Human-centered computing** →**Collaborative and social computing;** •**Mathematics of computing** →*Random graphs;* •**Applied computing** →*Sociology;*

## KEYWORDS

Network Diversity; Embeddedness; Network Signature; Triadic Closure; Link Prediction; Social Networks; Big Data.

---

*This work was done when Yuxiao was a P.h.D. student at University of Notre Dame.

## 1 INTRODUCTION

Since the time of Aristotle, it has been observed that people "love those who are like themselves" [2]. We now know this as the principle of homophily, which asserts that people's propensity to associate and bond with others similar to themselves is what drives the formation of social relationships [19, 27]. But homophily applies not only to shared traits and characteristics, as captured by this principle, but to the network structure of our relationships as well.

Structural homophily holds that individuals with more friends in common are more likely to associate [15, 30]. This principle has been widely explored in network science [1, 24], where it has been shown to be a strong driving force of link formation. However, structural homophily can be limiting in its explanation of the diverse ways that may actually drive the process of link formation. While it accounts for similarity based on the number of common neighbors, structural homophily fails to account for the diverse ways in which these neighbors are embedded within the network.

We posit that structural diversity, as measured by the number of connected components [39], may provide additional insight into how individuals are connected. For example, common social network neighbors may arise from relatively weak connections such as a shared hometown, implying a more diverse common neighborhood, or relatively stronger connections such as being college classmates, implying a relatively less diverse common neighborhood. In this paper, we study this notion of *structural diversity of the common neighborhood shared by two nodes*, investigating how it aligns with the principle of homophily and whether it can shed light on the process of link formation and network organization.

**Motivating example.** Consider the scenarios presented in Figure 1. According to homophily, the probability that $v_i$ and $v_j$ know each other given that they share four (b) common neighbors (CN), is generally higher than when they share only three (a). Formally, $P(e_{ij}=1 \mid \#CN=4) > P(e_{ij}=1 \mid \#CN=3)$, where $e_{ij}=1$ denotes the existence of an edge $e$ between $v_i$ and $v_j$. A natural question that arises is how two users' common neighborhood—that is, the subgraph structure of their common neighbors—influences the probability that they form a link. For example, let us assume that $v_i$ and $v_j$ share four common neighbors. Are $v_i$ and $v_j$ more likely to connect with each other if their four common neighbors do not know each other (i.e., (b) ▨), or if they all know each other (i.e., (c) ▨)? In essence, then, we are interested in the truth of the following inequality:

$$P(e_{ij}{=}1 \mid \text{▨}) \; \geqslant \; P(e_{ij}{=}1 \mid \text{▨}) \; ?$$

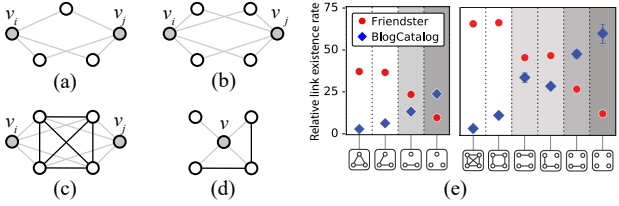

**Figure 1: Structural diversity of common neighborhoods.** Two nodes $v_i$ and $v_j$ with three disconnected (a), four disconnected (b), and four fully connected (c) common neighbors. (d) Different from structural diversity in an ego-centric notion [39], we go beyond an ego, and focus on the structural diversity of common neighborhoods for two persons. (e) Structural diversity of common neighborhoods affects the link existence rate.

**Contribution.** We study the structural diversity of common neighborhoods between two nodes and its impact on the probability that these nodes may be connected. The diversity measures the number of connected components that comprise the common neighborhood, capturing the variable ways in which a given pair's common contexts may be composed. Our study considers more than one hundred large-scale networks from a wide range of domains, *making this the largest empirical analysis done on networks to date*. We first consider three representative networks—Friendster, BlogCatalog, and YouTube—to develop the intuition underlying these ideas, and demonstrate the notion of common neighborhood diversity and its impact on link existence.

We discover, in Figure 1(e), that when the size of the common neighborhood is fixed with the same diversity (i.e., #components), the link existence rates given different common neighborhoods (e.g., ▨ vs. ▨, ▨ vs. ▨, and ▨ vs. ▨) are relatively close. Further, we find an increase in the structural diversity of the neighborhood negatively impacts the formation of links in Friendster—that is, $P(e{=}1 \mid \text{▨}) > P(e{=}1 \mid \text{▨})$—while it actually facilitates the formation of links in BlogCatalog—that is, $P(e{=}1 \mid \text{▨}) < P(e{=}1 \mid \text{▨})$.

We also discover striking phenomena where structural diversity violates the principle of homophily. When applied to the context of common neighborhoods, the principle suggests, for example, that $P(e{=}1 \mid \#CN{=}4) > P(e{=}1 \mid \#CN{=}3)$. However, if we consider BlogCatalog, for example, we find that the link existence rate of four common neighbors in a single component is significantly lower than the rate of only three disconnected common neighbors, i.e., $P(e{=}1 \mid \text{▨}) < P(e{=}1 \mid \text{▨})$. Similarly, this violation is also observed in Friendster, i.e., $P(e{=}1 \mid \text{▨}) < P(e{=}1 \mid \text{▨})$.

We apply our discoveries to develop the common neighborhood signature (CNS). By leveraging the CNS of each network, we are able to cluster all 80 real-world networks used in our study into three unique superfamilies. Friendster, BlogCatalog, and YouTube are each representative of a particular superfamily. We find that these superfamilies cannot be uncovered by subgraph frequencies [28] and other properties, indicating that the CNS can serve to reveal previously undiscovered mechanisms of network organization. In addition, we find that classical random graph models are unable to generate synthetic graphs that have CNS similar to real networks, despite their degree distribution, clustering coefficient, and other traditional network properties matching those of real networks.

Investigating the cause of these differences across superfamilies, we find that they correspond to the ways in which these networks are used to satisfy social or information needs. Some networks, such as Friendster and Facebook, are more likely to be used for maintaining densely connected real-world relationships, resulting in such networks exhibiting a positive relationship between the diversity of the common neighborhood and the link existence rate. By contrast, other networks, such as BlogCatalog and LinkedIn, are more likely to be used for acquiring new and diverse information, resulting in a negative relationship between the common neighborhood diversity and the link existence rate.

Our findings demonstrate that the structure of common neighborhoods can effectively capture driving forces intrinsic to the formation of local network structures and global network organizations. The results have important, practical implications for random graph model design, as well as recommendations in social networks, such as "People You May Know (PYMK)" in Facebook. For example, by simply using structural diversity as an unsupervised link predictor, our case study finds a 57% improvement over the commonly used #CN predictor as measured by AUPR in both the Friendster and BlogCatalog Networks.

## 2 BIG NETWORK DATA

To comprehensively examine our proposed concept of the structural diversity of common neighborhoods, we have assembled a large collection of big network datasets from several platforms, including (in alphabetical order): AMiner [37], ASU [45], KONECT [18], MPI-SWS [29], ND [4, 6], NetRep [35], Newman [31], and SNAP [36].

In total, we have compiled a set of 120 large-scale undirected and unweighted networks from the platforms listed above, including 80 real-world networks and 40 random graphs. We have cleaned the networks as follows: For directed networks that have no reciprocal connections (5 citation networks), we convert each directed link into an undirected one. We prune the resulting undirected networks by retaining only the largest connected component.

The 80 real-world networks used in this work are shown in Table 4 (See the last page)[1], which provides detailed network characteristics. We note that the largest network used in this work is the *soc-Friendster-SNAP* online social network, which consists of over 65 million nodes and more than 1.8 billion edges.

We also test the concept of structural diversity of common neighborhoods on 40 random graphs, including 10 networks generated by each of the four following models: Erdős-Rényi (ER) [9], Barabási-Albert (BA) [4], Watts-Strogatz (WS) [43], and Kronecker [21]. For the first three models, the number of nodes is set as 1,000,000. In the ER model, we set the edge creation probability to between $5{\times}10^{-6}$ and $5{\times}10^{-5}$ with a step of $5{\times}10^{-6}$, thereby generating 10 ER random graphs with the number of edges ranging from roughly 2,000,000 to 25,000,000. We generate 10 BA random graphs with between 2,000,000 and 20,000,000 edges. We generate 10 WS random graphs by setting different mean degrees $k$ and rewiring probabilities $\beta$, where $k$ is chosen from 8, 12, 16, 20, and 24, and $\beta$ is 0.2 or 0.8. There are between 2,000,000 and 14,000,000 edges in the WS graphs. Finally, we use 10 Kronecker graphs with the original parameters (estimated by [21]) fitted to 10 real networks, which are among the 80 we use above. This means that we would expect the results discovered from the 10 Kronecker graphs to be in close agreement with the corresponding 10 real networks if the Kronecker model was capable of preserving the common neighborhood in addition to traditional network properties, such as degree, triangle, and diameter distributions.

---

[1] For detailed information about the 120 networks and source code, please refer to the companion webpage at https://ericdongyx.github.io/diversity/cns.html.

# 3 COMMON NEIGHBORHOOD DIVERSITY

In this section, we define basic concepts and formalize the problem. Formally, we use $G = (V, E)$ to denote an undirected and unweighted network, where $V = \{v_i\}$ represents the set of nodes and $E \subseteq V \times V$ represents the set of links between two nodes. We denote each existing link, $e_{ij} \in E$, as $e_{ij} = (v_i, v_j) = 1$ and each non-existing link, $e_{ij} \notin E$, as $e_{ij} = (v_i, v_j) = 0$.

*Common Neighborhood:* Let $N(v_i)$ denote the adjacency list of a node $v_i$, i.e., $v_i$'s neighborhood. The common neighborhood of each pair of two nodes $v_i$ and $v_j$ can be represented as the subgraph composed of their common neighbors, $G^{ij} = (V^{ij}, E^{ij})$, where $V^{ij} = N(v_i) \cap N(v_j)$ denotes the common neighbors of $v_i$ and $v_j$, and $E^{ij} = \{e_{pq} \mid e_{pq} \in E, v_p \in V^{ij}, v_q \in V^{ij}\}$ denotes the edges among their common neighbors.

**Input:** Given a network $G = (V, E)$, the input of our problem includes (1) each pair of users who have at least one common neighbor, i.e., $\{(v_i, v_j) = e_{ij} \mid |V^{ij}| \geq 1\}$, and (2) the common neighborhood $G^{ij} = (V^{ij}, E^{ij})$ of each pair of users $v_i$ and $v_j$.

*Structural homophily:* The principle of structural homophily suggests that with more common neighbors, it is more likely for two people to know each other. Formally, this means that if $y > x$, then $P(e_{ij}=1 \mid |V^{ij}|=y)$ should generally be greater than $P(e_{ij}=1 \mid |V^{ij}|=x)$. A long line of work from various fields has demonstrated that this principle holds across a wide variety of different networks.

In this study, we revisit this principle of structural homophily, further proposing to study the structure of common neighborhoods. We characterize the structure of a small graph as its diversity, formalizing its application to common neighborhoods below.

*Definition 3.1.* **Structural Diversity of Common Neighborhoods:** Given a network, $G$, a pair of users in this network, $v_i$ and $v_j$, and the pair's common neighborhood, $G^{ij} = (V^{ij}, E^{ij})$, we define the structural diversity of the common neighborhood as the number of connected components in $G^{ij}$, denoted $|C(G^{ij})|$.

Consider a pair of users with four common neighbors. This pair's common neighborhood has 11 possible configuration structures: 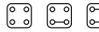. In general, we refer to the more diverse structure as the one with more components.

**Output:** Our goal is to study the relations between the structure of two users' common neighborhood and the probability that there exists a link between these two users. Formally, this means that given two users $v_i$ and $v_j$ and their common neighborhood $G^{ij}$, the output of our problem is the link existence probability distribution of $G^{ij}$, denoted $P(e_{ij}=1 \mid G^{ij})$.

In each network, we can enumerate all pairs of users who have at least one common neighbor. If we fix the common neighborhood $G^{ij}$ of $v_i$ and $v_j$, then we can compute the link existence probability $P(e_{ij}=1 \mid G^{ij})$ based on the number of link pairs that exist.

*Relative Link Existence Rate:* To facilitate the comparability of results across networks with diverse sizes and densities, we define the relative link existence rate $R(e_{ij}=1 \mid G^{ij})$ as

$$R(e_{ij}=1 \mid G^{ij}) = P(e_{ij}=1 \mid G^{ij}) \ / \ P(e=1 \mid \#CN=1),$$

where $P(e=1 \mid \#CN=1)$ denotes the link existence probability when two users have exactly one common neighbor.

*Definition 3.2.* **Common Neighborhood Signature (CNS):** Given a network $G = (V, E)$, its common neighborhood signature is defined as a vector of relative link existence rates with respect to the specified common neighborhoods. Each element of this vector is a relative link existence rate corresponding to a particular common neighborhood structure.

Consider, for example, user pairs with between two and four common neighbors. The common neighborhoods that can be represented by the pairs of users with two-to-four common neighbors correspond to a vector of the relative link existence rates for 17 subgraphs: 2 for common neighborhoods with size two, 4 for those with size three, and 11 for those with size four.

Given this input and output, our work seeks to understand the underlying driving forces behind link formation and network organization by answering the following questions:

- How does the structural diversity of common neighborhoods influence the link existence probability?
- Does structural diversity concord or conflict with the principle of homophily in networks?
- Does the common neighborhood signature (CNS) vary across different networks?
- How does the common neighborhood signature differ from subgraph frequency?

# 4 DIVERSITY IN LINK EXISTENCE

To understand how the structural diversity of common neighborhoods influences link existence, we investigate three representative networks in Table 4: Friendster, BlogCatalog, and YouTube.

## 4.1 How Does Diversity Affect Link Existence?

When we control for the size of the common neighborhood between two users, how does the structure of the common neighborhood influence the probability that the pair forms a link in the network? An illustrative example of this question is introduced as follows: Given that two users $v_i$ and $v_j$ have four common neighbors, are they more likely to connect with each other if their four common neighbors do not know each other ( 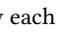 ) or if their four common neighbors already know each other ( 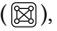 ), i.e.,

$$P(e_{ij}=1 \mid \text{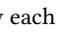}) \ \geqslant \ P(e_{ij}=1 \mid \text{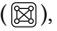}) \ ?$$

To address this question, we compute all 546 billion, 612 million, and 1.26 billion pairs of users who have at least one common neighbor in the Friendster, BlogCatalog, and YouTube networks, respectively. Figure 2 presents the relative link existence rates in the three networks for neighborhoods with between two and six common neighbors. It is immediately observable that the impact of structural diversity on link existence in the three networks is remarkably varied.

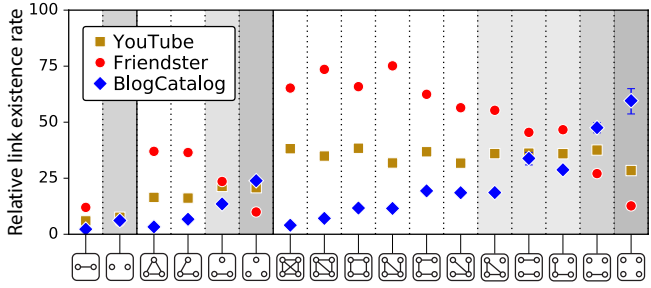
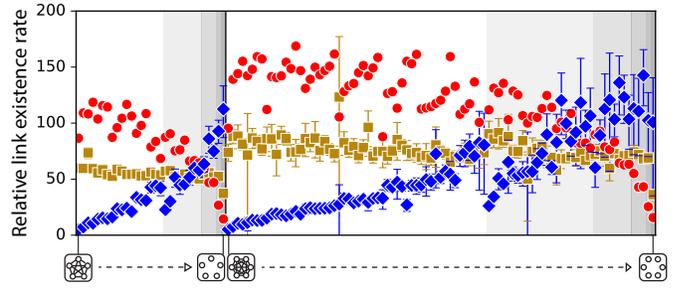

**Figure 2: Structural diversity of common neighborhoods in link existence.** *x*-axis: two-node, three-node, and four-node common neighborhoods on the left side; five-node and six-node common neighborhoods on the right side. The *x*-axis is ordered according to the following keys: common neighborhood size (ascending), component count of the common neighborhood (ascending), and edge density of the common neighborhood (ascending). When all three keys are the same, the degree sequence of the common neighborhood is in descending order. Shading indicates differences in the number of components. Error bars designate the 95% confidence interval.

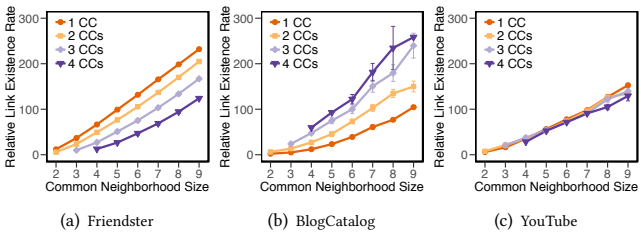

**Figure 3: Structural diversity vs. link existence.** Relative link existence rate as a function of component count (diversity).

Recall that diversity is measured by the number of connected components. In general, if we control for the size of the common neighborhood, then as the component count increases—as distinguished by different shadings—the link existence rate decreases in Friendster but increases in BlogCatalog. For example, consider the cases where the pair have exactly four common neighbors. If we consider the common neighborhoods ordered according to increasing diversity (more components), we have: (1 component) ⊠ ⊠ ⊠ ⊠ ⊠ ⊠ → (2) ⊠ ⊠ ⊠ → (3) ⊠ ⊠ → (4) ⊠. As we move from left to right in the ordering (corresponding to increasing diversity), it is less likely for two people in Friendster to associate with each other—also confirmed in Figure 3. This tells us that users on BlogCatalog are more likely to connect if their common friends are more structurally diverse, while users on Friendster are more likely to connect if their common friends are more densely embedded within the same community.

We also quantify the impact of diversity on link existence. Table 1 reports the Pearson correlation coefficients $\rho$ between the relative link existence rate and the structural diversity of common neighborhoods. We can clearly see that structural diversity is strongly correlated ($|\rho| > 0.8$) with the link existence rate in Friendster and BlogCatalog, but with opposing signs.

In addition, for common neighborhoods with the same size (fixed homophily) and component count (fixed diversity), we still observe variations in the relative link existence rate. When both the common neighborhood size and its component count is controlled for, the link existence rates vary as a function of edge count within its components. For example, when all four common neighbors of two users belong to one component in BlogCatalog, the probability for this pair to connect consistently decreases as the edge count increases—(3 edges) ⊠ → (4) ⊠ (5) → ⊠ → (6) ⊠, while they have similar probabilities if their common neighborhood has the same number of edges, i.e. (three edges) ⊠ and ⊠, and (four edges) ⊠ and ⊠. Indeed, edge count in the common neighborhood also has a crucial role in determining link existence when homophily and diversity are both fixed.

**Table 1: Correlation analysis for relative link existence rate.**

| #CN | 2 | 3 | 4 | 5 | 6 |
|---|---|---|---|---|---|
| Friendster | −1.0 | −0.99 | −0.95 | −0.88 | −0.79 |
| BlogCatalog | 1.0 | 0.98 | 0.94 | 0.89 | 0.67 |
| YouTube | 1.0 | 0.86 | −0.38 | −0.55 | −0.38 |

**Summary.** We demonstrate that the structural diversity of common neighborhoods is a crucial factor in determining link existence across networks. If we fix the homophily effect (#common-neighbors), we find that the structural diversity of common neighborhoods has a negative effect on the formation of online friendships in Friendster but a positive effect in BlogCatalog, and a relatively neutral effect on YouTube. This reveals a fundamental difference between these three networks in their microscopic structures and link formation mechanisms.

### 4.2 Does Diversity Violate Homophily?

Previously, we demonstrated that in Friendster the structural diversity of common neighborhoods is in general negatively associated with link existence, i.e., $P(e{=}1|\includegraphics{}) < P(e{=}1|\includegraphics{})$, while in BlogCatalog it is in general positively associated with link existence, i.e., $P(e{=}1|\includegraphics{}) > P(e{=}1|\includegraphics{})$. A subsequent question one may ask is whether structural diversity conflicts with the principle of homophily. Specifically, we can formalize this question for Friendster and BlogCatalog in the following way:

$$P(e{=}1|\includegraphics{}) > P(e{=}1|\includegraphics{})? \quad and \quad P(e{=}1|\includegraphics{}) > P(e{=}1|\includegraphics{}) ?$$

Conventional wisdom may answer "yes" to both cases, as the concept of structural homophily suggests that, all other things being equal, relationships are more likely to form between individuals that share a larger common neighborhood, which is further empirically demonstrated in Figure 4(a). Surprisingly, however, we find that there is no empirical evidence to support the existence of homophily within the context of structural diversity.

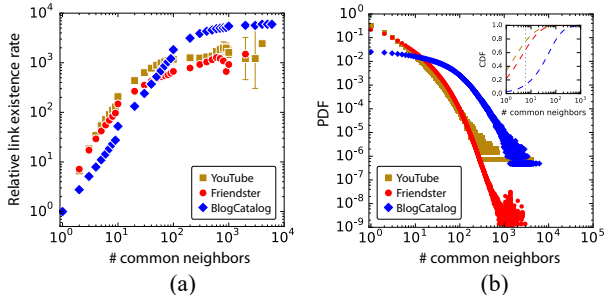

(a)                                    (b)

**Figure 4: Common neighbor characterization.** (a) The link existence rate as a function of #common neighbors. (b) The probability density function (PDF) of #common neighbors.

In Figure 2, we can observe that the link existence rate between two BlogCatalog users with less diversely connected common neighbors is actually *lower* than the link existence rate between users with fewer but more diversely connected neighbors. For example, if two users share four common neighbors, the probability that there exists a link between them is, in more than half of the eleven configurations (⊠ ⊠ ⊠ ⊠ ⊡ ⊞), lower than the probability of a link between users that share three disconnected common neighbors (⊡). In fact, $P(e=1|⊡)$ is 493% higher than $P(e=1|⊠)$; even $P(e=1|⊡)$ is higher than $P(e=1|⊠)$. By contrast, in Friendster, homophily is instead violated when a larger number of disjoint common neighbors meets with a smaller number of connected ones. For example, the link existence probabilities given ⊡ and ⊡ are lower than those given ⊠ and ⊡. Similar violations occur in various cases with different common neighborhood sizes.

**Summary.** Our observations show that structural diversity does indeed, in many cases, violate the principle of homophily. These observations suggest that the fundamental assumption held by the homophily principle that "more common friends means a higher probability to connect" can often be an oversimplification, and—as we clearly show—is not necessarily true.

## 5  CNS FOR NETWORK SUPERFAMILIES

The global structure of network systems are governed by natural laws, through which constant, universal properties such as long-tailed degree distributions arise. However, even when subject to the global structures prescribed by these laws, different networks can still reveal distinct local properties and structures. One striking example is the discovery that networks with long-tailed degree distributions can be naturally cataloged into distinct superfamilies of networks based on their subgraph frequencies [28]. In this section, we investigate how the common neighborhood signature can—like subgraph frequency—uncover previously undiscovered mechanisms of network organization, thereby allowing it to serve as a unique property by which to categorize networks.

### 5.1  Can CNS Identify Network Superfamilies?

To answer this question, we examine the similarity between the functions of structural diversity of common neighborhoods across the 120 real-world and random networks studied. We begin by constructing, for each network, the common neighborhood signature represented by neighborhoods with between two and four common neighbors (constituting a length-17 vector). We then use this signature to compute the similarity in the structural diversity for each pair of networks.

Figure 5 visualizes the similarity matrix of the common neighborhood signatures between every pair of networks. The $x$- and $y$-axes represent all 120 networks studied in this work, and the spectrum color represents the correlation coefficient. Note that the arrangement of rows and columns in the presented similarity matrix is determined by the Ward variance minimization algorithm for hierarchical clustering [42], and the similarity between common neighborhood signatures is measured by the Pearson correlation coefficient [28].

**Network Superfamilies.** We observe three major clusters on the top right of Figure 5(a), which mostly consist of real networks. There exist several minor clusters on the lower left, which corresponds to most of the random graphs. The fact that our similarity analysis distinguishes between real and artificial networks suggests that the common neighborhood signatures capture hidden properties underlying the network structures.

According to the common neighborhood signature, a total of 36 networks—including Friendster and Facebook—are grouped together into a single cluster (colored 'red' in the dendrogram), in which the structural diversity of common neighborhoods has similar effects on link formation in each network. Another 29 networks—including BlogCatalog and LinkedIn—are grouped into another cluster (colored 'blue'), indicating strong correlations among these networks but weak correlations between these networks and those in the red cluster. The overlap of the two clusters, if considered separately, consists of 24 additional networks (colored 'gold'). We note that the networks in the gold cluster (the overlapping part) demonstrate relatively higher similarity with networks in the red and blue clusters than the networks in the red and blue clusters demonstrate with each other. Finally, there are 31 remaining networks (largely comprising of random graphs) that are not clustered into any of the three aforementioned clusters (colored 'black').

To summarize our key findings of the three families, and their representative characteristics to help explain the aspects of homophily and structural diversity:

- Networks belonging to the 'red' family, such as Friendster and Facebook, are more likely to be used for maintaining friendships that can be recognized in the real world, leading to densely connected local groups. For example, students from the same academic department have the propensity to be connected as Facebook friends. As a result, in these networks we observe that common neighborhood diversity has a negative effect on the relative link existence rate—that is, as the diversity (heterogeneity) of a pairs' common neighborhood decreases, the probability of a connection between the pair actually increases.
- Networks belonging to the 'blue' family, such as BlogCatalog and LinkedIn, are typically used for acquiring information and

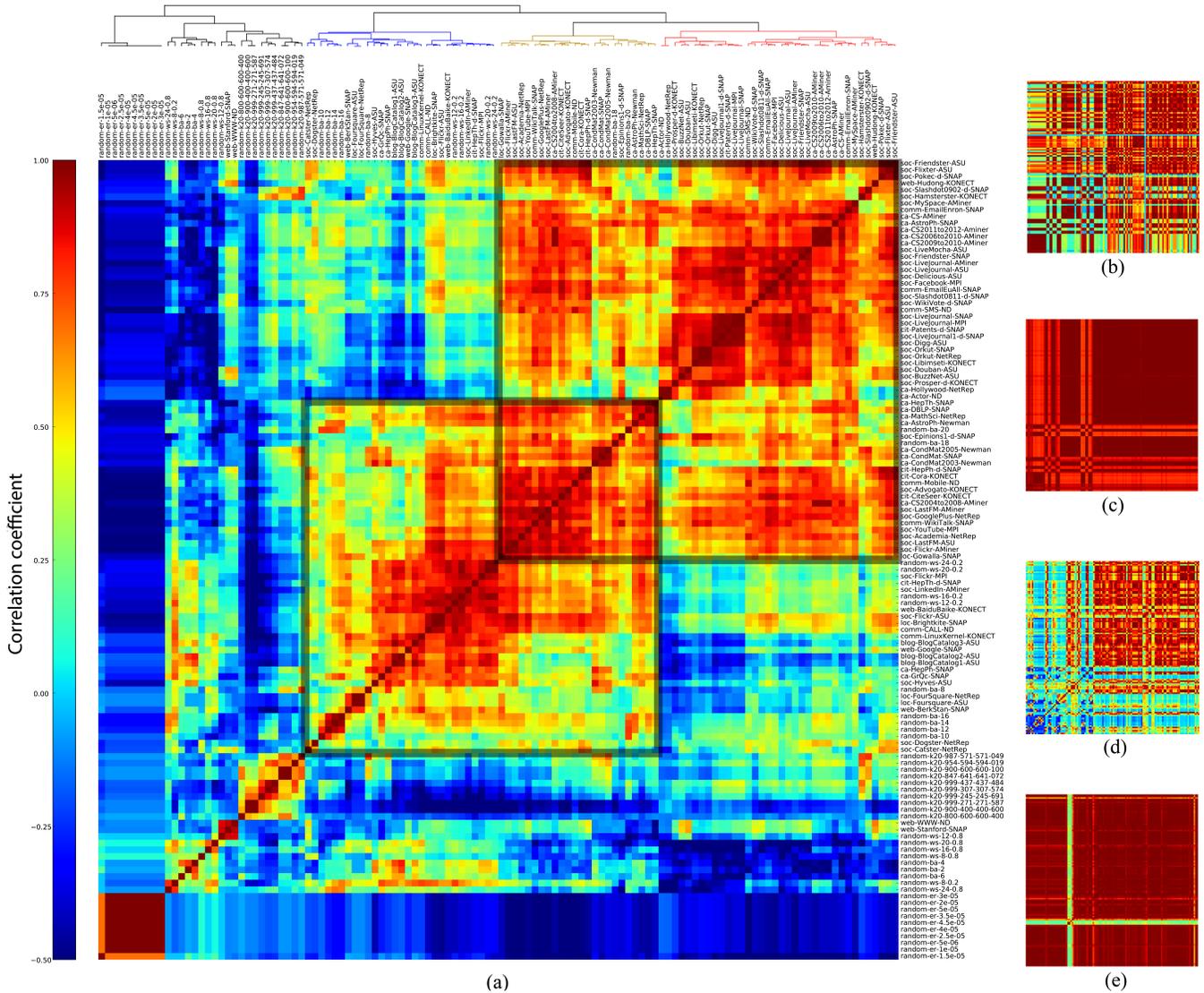

**Figure 5: Correlation coefficient matrix of different methods for 120 networks.** (a) Common neighborhood signature. (b) Subgraph significance profile. (c) Sequence of percentile degrees. (d) Bag of degrees. (e) Bag of #CNs.

resources, motivating people to strategically diversify their connections for the purpose of obtaining more information sources. For example, it has been shown that in LinkedIn, weak ties—relationships that are not densely embedded—are the source of a majority of job opportunities provided through the network[2]. Consequently, in these networks, as the diversity (heterogeneity) of a pairs' common neighborhood increases, the probability of a connection between the pair also increases.

- Networks belonging to the 'gold' family are a blend of 'blue' and 'red' families, including YouTube and many author collaboration networks. If we consider the latter as an example, there is an expectation of maintaining close collaborations between two scholars and their coauthors; at the same time, there is also an expectation of the expansion of a scholar's access to new (diverse) collaborations, resulting in a neutral relationship between the common neighborhood diversity and link formation.

**Implications for Random Graph Models.** We further study how the common neighborhood signatures qualify the nature of random graphs. Observed from Figure 5(a), all Erdős-Rényi (ER) graphs [9] are densely clustered into the bottom-left hierarchy. According to Watts and Strogatz [43], WS random graphs with the $\beta$ parameter close to 1 tend to approach ER random graphs. This theory is captured by the structural diversity profile, as WS graphs with $\beta$=0.8 show the highest similarity with ER graphs.

We find it striking that while WS and BA graphs may imitate specific superfamilies (blue or gold) of real networks (e.g., BlogCatalog and LinkedIn), none of them are able to simulate an important family (red) of real-world networks (e.g., Facebook and Friendster).

---

[2] https://www.fastcompany.com/3060887/the-future-of-work/what-linkedin-data-reveals-about-who-will-help-you-get-your-next-job

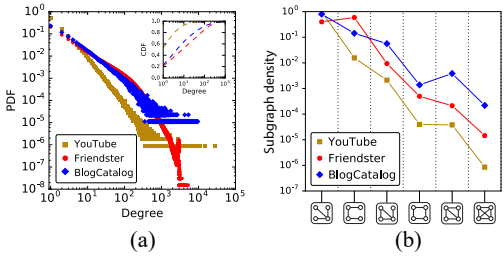

**Figure 6: Degree distribution (a) and four-node subgraph frequency distribution (b) in three networks.**

Even the 10 Kronecker graphs with the original parameters fitted to 10 real networks do not belong to any of the three major superfamilies, but instead form a standalone cluster. These observations indicate that although the random graph models studied may satisfy a series of network properties, the common neighborhood signature is a new network property that is not adequately captured by these models. *To this end, our results have important implications for random graph model design,* suggesting that in addition to fitting traditional network properties [21, 22, 34] such as degree distribution, clustering coefficient, associativity, diameter, etc., random generation models that leverage CNS may lead to graphs that are more representative of real-world networks.

### 5.2 CNS vs. Subgraph Frequency?

Characterization of the global similarity and difference across multiple networks is conventionally focused on degree distribution [4, 11], degree sequence [10, 32], and subgraph frequency [28]. To examine the significance of the common neighborhood signature, we need to investigate not only its ability to effectively characterize networks, but the extent to which these characterizations are distinct from those provided by these conventional methods.

As there are several ways to quantify network properties at the global scale, we compare the common neighborhood signature with the following four conventional approaches: The subgraph significance profile [28]; the degree sequence vector [10]; the bag-of-degrees vector (degree distribution); and the bag-of-#CNs vector proposed in this work (Figure 4(b)).

**Micro.** We analyze the CNS over three representative networks—Friendster, BlogCatalog, and YouTube. However, our findings generally hold for any network within a given superfamily. At the micro scale, we can examine the visualized distributions of the aforementioned measures. The four-subgraph distributions (computed by ESCAPE [33]) are shown in Figure 6(b), the degree distributions in Figure 6(a), and the common neighborhood size distributions in Figure 4(b). From these figures, we can observe that, while not identical, each type of distribution reveals similar trends and shapes within the three networks. We also provide numerical results of the differences between Friendster and BlogCatalog. The correlation coefficients based on subgraph, degree sequence, bag of degrees, and bag of #CNs are 0.579, 1.000, 0.996, and 0.957, respectively, which are significantly higher than common neighborhood signature-based quantification ($-0.267$). *The strong correlations between Friendster and BlogCatalog produced by all four alternative methods further highlight their inability to uncover hidden network properties.*

**Table 2: Regression analysis for relative link existence rate.** Significance code: $^{*}\ p < 0.05$; $^{**}\ p < 0.01$; $^{***}\ p < 0.001$.

| Network | Friendster | BlogCatalog | YouTube |
|---|---|---|---|
| Intercept | $-0.03845$ *** | 0.00010 | $-0.01855$ *** |
| Homophily (#CN) | 0.01948 *** | 0.00252 *** | 0.00792 *** |
| Diversity (#components) | $-0.01102$ *** | 0.00114 *** | $-0.00047$ |
| Adj. $R^2$ (Diversity) | 0.83330 | 0.76750 | 0.81440 |
| Adj. $R^2$ (Homophily) | 0.42300 | 0.14260 | 0.77160 |

**Macro.** At the macro scale, we can examine heatmaps of the correlation coefficient matrix for each method in Figure 5 for all 120 networks. For comparison with the CNS, the ordering of networks in these four matrices are kept identical to that in Figure 5(a)[3]. The four resulting matrices fail to show clear and dense clusters, further supporting the ability of common neighborhood signatures to detect unique network superfamilies. Based on these results, we argue that *the common neighborhood signature is able to capture underlying mechanisms of network organization that cannot be discovered by conventional methods such as the subgraph significance profile [28], degree distribution [11], and degree sequence [32].*

**Summary.** Our comprehensive study based on both micro- and macro-level phenomena demonstrates that the CNS can detect intrinsic, hidden network superfamilies that are not discoverable by conventional methods. The difference between uncovered superfamilies lie in the distinct strategies that people use across different networking services for satisfying various needs, such as the use of Friendster ('red' family) for satisfying social needs and BlogCatalog ('blue' family) for satisfying information needs. Together with classical network properties, we also find that CNS can be used to examine the fitness of random graphs in simulating real networks.

## 6 A USE CASE: LINK INFERENCE

We posit that the structural diversity of common neighborhoods is crucial in determining link existence, more so than the principle of structural homophily. More formally, we ask which of the following measurements is more accurate,

$$P(e=1\mid\boxtimes)\ \cdots\ P(e=1\mid\boxdot)\quad\text{or}\quad P(e=1\mid\text{\#CN}{=}4)\ ?$$

First, to demonstrate the role of structural diversity in link inference, we perform a regression analysis with the relative link existence rate serving the dependent variable, shown in Table 2. We find that structural diversity can be used as a highly accurate predictor of the link existence rate ($R^2 > 0.76$). We also find that structural diversity serves as a statistically significant ($p < 0.001$) link prediction factor in the Friendster and BlogCatalog networks. Observed from the last row of Table 2, when predicting for Friendster and BlogCatalog, we can achieve a far better estimation by using the structural diversity of common neighborhoods than from using only the number of common neighbors, as measured by $R^2$. On Friendster, the $R^2$ improves from 0.42 to 0.83 (+97%), and on BlogCatalog, the $R^2$ improves from 0.14 to 0.76 (+442%).

Second, we use both structural homophily and diversity as link predictors to infer whether there exists a link between two users.

---

[3]Note that the subgraph significance profile (Figure 5(b)) is able to categorize the networks into different superfamilies if the same clustering algorithm is applied to the correlation matrix. However, the goal here is to see whether CNS can offer new and meaningful perspectives for network superfamily, rather than its advantage over others.

**Table 3: Link prediction by #CN and diversity.**

| Metric | Method | Friendster | BlogCatalog | YouTube |
|--------|--------|-----------|-------------|---------|
| Data | #Pairs | 67,033,108,105 | 224,786,028 | 118,635,122 |
| | %Positive | 0.91830% | 0.09430% | 0.50820% |
| AUPR | Homophily | 0.02230 | 0.00178 | 0.01524 |
| | Diversity | 0.03499 | 0.00279 | 0.01532 |
| AUROC | Homophily | 0.68539 | 0.66259 | 0.69371 |
| | Diversity | 0.71722 | 0.70239 | 0.68401 |

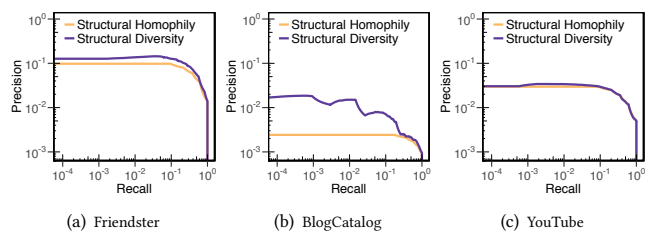

(a) Friendster    (b) BlogCatalog    (c) YouTube

**Figure 7: Precision-recall curves for inference of link existence by #CN and diversity.**

For these predictions, we limit the candidate pairs of users to be inferred as those users with between two and six common neighbors. This generates more than 67 billion, 224 million, and 118 million data instances in the Friendster, BlogCatalog, and YouTube networks, respectively. We further note that the ratio between positive instances (existing links) and negative instances (non-existing links) is highly skewed in each network, complicating the prediction task.

Table 3 shows the link prediction performance generated by structural homophily and diversity on each of the three networks as measured by AUPR and AUROC. Figure 7 illustrates the corresponding precision-recall curves. In terms of AUPR, the structural diversity-based unsupervised predictor outperforms the homophily-based predictor by about 57% in the Friendster and BlogCatalog networks. In terms of AUROC, structural diversity also demonstrates greater predictive power than homophily. A $t$-test finds that the improvement of the diversity-based predictor over the homophily-based predictor is highly statistically significant ($p \ll 0.001$).

Note that the structural diversity-based predictor does not outperform the structural homophily-based predictor on the YouTube network. While the lack of improvement could be considered disappointing, this result actually further validates the findings in Figure 2, which shows that the impact of the structural diversity of common neighborhoods on link existence can differ among networks of different superfamilies. Specifically, this result shows that the influence of structural diversity on networks in the 'gold' superfamily, which includes the YouTube network, is not as significant as it is for the 'blue' and 'red' superfamilies. That is, the observed difference in performance is a consequence of the underlying factors that distinguish the 'gold' superfamily from the others.

**Summary.** We provide empirical evidence that the structural diversity of common neighborhoods helps the link inference task for networks in the 'blue' and 'red' superfamilies, and we demonstrate that this performance reaffirms the existence of superfamilies. We find that proper application of structural diversity has the potential to substantially improve the predictability of link existence, with important implications for improving recommendation functions employed by social networking sites and guiding our understanding about the link formation processes.

## 7 RELATED WORK

Social theories are the empirical abstraction and interpretation of social phenomena at a societal scale. The idea of homophily, in particular, dates back thousands of years to Aristotle, who observed that people "love those who are like themselves" [2]. In this modern-day conception, the principle of homophily holds that individuals are more likely to associate and bond with similar others [19, 27]. In the context of network science, structural homophily suggests that people with more common neighbors tend to connect with each other [15, 30].

**Embeddedness.** Granovetter defined embeddedness as "the extent that a dyad's mutual contacts are connected to one another" [13, 14], followed by a line of work that has demonstrated the power of embeddedness in network organization and economic development [41]. Note that sometimes this concept can be also considered to be "the number of common neighbors the two endpoints have" [8, 26]. But even with a fixed number of edges between a dyad's common neighbors, there still exist a diverse set structures that describe common neighborhoods [3, 5]. How these diverse structures influence link existence remains an open question in both network and social science.

**Structural Diversity.** The concept of structural diversity was first proposed in an ego network by Ugander et al. [39], who found that the user recruitment rate in Facebook is determined by the variety of an individual's contact neighborhood, rather than the size of his or her neighborhood. Further studies show that the diversity of one's ego network also has significant influence on a user's other social decisions [12, 25]. The difference between this work and our study centers around neighborhood studies. While Ugander et al. study the variety of a single individual's contact neighborhood, we instead focus on the structural diversity of a pair of individual's common neighborhoods.

**Subgraph and Network Superfamily.** Milo et al. investigated the distributions of subgraph frequency across multiple types of networks, and proposed a subgraph-based significance profile for networks [28]. By leveraging this profile, they discovered several network superfamilies, whereby networks in the same superfamily display similar subgraph distributions. Our work, however, is different from subgraph mining [28, 38, 38], graph classification [16], and the graph isomorphism problem [40]. Instead, our work focuses on uncovering the principles that drive the formation of local network structure and exploring the significance of structural diversity in driving link organization and network superfamily detection.

The structural diversity of common neighborhoods may also offer substantial potential for applications to other important network mining tasks, such as link prediction [7, 24], social recommendation [25], tie strength modeling [3, 44], community detection [23], and network evolution [17, 20, 22].

## 8 CONCLUSION

In this work, we study how the different common neighborhood configurations can influence network structures. Through a comprehensive study of 120 big real-world and random networks, we conclude that, after controlling for the number of common neighbors, the structural diversity of common neighborhoods exhibits substantial influence on link existence rates. We further define the common neighborhood signature (CNS) of a network, which classifies the 120 networks into three unique superfamilies that we find are not discoverable by conventional network properties.

Strikingly, we find that LinkedIn and Facebook belong to different network superfamilies. Characterizing their distinctions, we find, for example, that when the size of common neighborhood is fixed to 3, an increase in its embeddedness actually negatively impacts the formation of professional relationships in LinkedIn and links in its superfamily (i.e., $P(e=1|\triangle) < P(e=1|\square)$), while it actually facilitates the formation of online friendships in Facebook and links in its superfamily (i.e., $P(e=1|\square) < P(e=1|\triangle)$). We conjecture that these distinctions arise from the different reasons for which these networks are used. These distinctions have important implications for "PYMK" functions used in both platforms, which also have an impact on link formation and network organization.

Furthermore, we find that none of the representative random graph models studied in this work are able to simulate a particular superfamily of real-world networks (e.g., Facebook and Friendster). Tackling this deficiency provides new opportunities and suggestions for building random graph models, which will also be one of our future research directions. One of our next steps will also be to extend our examination of common neighborhood structure beyond homogeneous and static networks, including heterogeneous networks and dynamic, inter-genre, attributed networks. Finally, we intend to incorporate the CNS into machine learning frameworks to improve social recommendation performance.

**Acknowledgments.** We would like to thank Omar Lizardo, Andrew Tomkins, and Zoltán Toroczkai for discussions. This work is supported by the Army Research Laboratory under Cooperative Agreement Number W911NF-09-2-0053 and the National Science Foundation (NSF) grants BCS-1229450 and IIS-1447795.

**Table 4: The statistics of 80 real networks.** Node degree and clustering coefficient are denoted by *d* and *cc*, respectively.

| Id | Network Name | #nodes | #edges | average d | 10% d | 50% d | 90% d | max d | average cc | global cc | diameter | #triangles | #triplets |
|---|---|---|---|---|---|---|---|---|---|---|---|---|---|
| 1 | **blog-BlogCatalog1-ASU** | 88,784 | 2,093,195 | 47.1525 | 1 | 5 | 88 | 9,444 | 0.3533 | 0.0624 | 9 | 51,193,389 | 2,410,854,351 |
| 2 | blog-BlogCatalog2-ASU | 97,884 | 1,668,647 | 34.0944 | 1 | 4 | 48 | 27,849 | 0.4921 | 0.0403 | 5 | 40,662,527 | 2,986,356,365 |
| 3 | blog-BlogCatalog3-ASU | 10,312 | 333,983 | 64.7756 | 4 | 21 | 136 | 3,992 | 0.4632 | 0.0973 | 5 | 5,608,664 | 167,281,662 |
| 4 | ca-Actor-ND | 374,511 | 15,014,839 | 80.1837 | 8 | 36 | 166 | 3,956 | 0.7788 | 0.1867 | 13 | 346,728,049 | 5,225,759,780 |
| 5 | ca-AstroPh-Newman | 14,845 | 119,652 | 16.1202 | 2 | 8 | 41 | 360 | 0.6696 | 0.5937 | 14 | 754,159 | 3,056,896 |
| 6 | ca-AstroPh-SNAP | 17,903 | 196,972 | 22.0044 | 2 | 10 | 55 | 504 | 0.6328 | 0.4032 | 14 | 1,350,014 | 8,694,840 |
| 7 | ca-CondMat2003-Newman | 27,519 | 116,181 | 8.4437 | 2 | 5 | 18 | 202 | 0.6546 | 0.3393 | 16 | 228,093 | 1,788,720 |
| 8 | ca-CondMat2005-Newman | 36,458 | 171,735 | 9.4210 | 2 | 5 | 20 | 278 | 0.6566 | 0.2903 | 18 | 374,300 | 3,493,465 |
| 9 | ca-CondMat-SNAP | 21,363 | 91,286 | 8.5462 | 2 | 5 | 18 | 279 | 0.6417 | 0.3172 | 15 | 171,051 | 1,446,763 |
| 10 | ca-CS2004to2008-AMiner | 434,357 | 1,578,275 | 7.2672 | 2 | 4 | 15 | 723 | 0.6684 | 0.3705 | 27 | 3,451,794 | 24,501,202 |
| 11 | ca-CS2006to2010-AMiner | 543,452 | 2,066,296 | 7.6043 | 2 | 4 | 16 | 1,082 | 0.6745 | 0.2939 | 25 | 4,372,725 | 40,256,430 |
| 12 | ca-CS2009to2010-AMiner | 315,263 | 1,059,740 | 6.7229 | 2 | 4 | 14 | 652 | 0.6989 | 0.4181 | 23 | 2,115,515 | 13,063,018 |
| 13 | ca-CS2011to2012-AMiner | 347,389 | 1,229,716 | 7.0798 | 2 | 4 | 14 | 793 | 0.7073 | 0.4004 | 29 | 2,585,990 | 16,787,160 |
| 14 | ca-CS-AMiner | 1,066,379 | 4,594,140 | 8.6163 | 1 | 4 | 18 | 1,983 | 0.6496 | 0.1873 | 24 | 10,312,677 | 154,825,662 |
| 15 | ca-DBLP-SNAP | 317,080 | 1,049,866 | 6.6221 | 1 | 4 | 14 | 343 | 0.6324 | 0.3850 | 23 | 2,224,385 | 15,107,734 |
| 16 | ca-GrQc-SNAP | 4,158 | 13,422 | 6.4560 | 1 | 3 | 15 | 81 | 0.5569 | 1.0829 | 17 | 47,779 | 84,582 |
| 17 | ca-HepPh-SNAP | 11,204 | 117,619 | 20.9959 | 2 | 5 | 47 | 491 | 0.6216 | 1.1768 | 13 | 3,357,890 | 5,202,255 |
| 18 | ca-HepTh-SNAP | 8,638 | 24,806 | 5.7435 | 1 | 3 | 13 | 65 | 0.4816 | 0.3460 | 18 | 27,869 | 213,790 |
| 19 | ca-Hollywood-NetRep | 1,069,126 | 56,306,653 | 105.3320 | 6 | 31 | 212 | 11,467 | 0.7664 | 0.3900 | 12 | 4,916,220,615 | 32,896,279,137 |
| 20 | ca-MathSci-NetRep | 332,689 | 820,644 | 4.9334 | 1 | 3 | 11 | 496 | 0.4104 | 0.1504 | 24 | 576,778 | 10,928,378 |
| 21 | cit-CiteSeer-KONECT | 365,154 | 1,721,981 | 9.4315 | 1 | 5 | 19 | 1,739 | 0.1832 | 0.0513 | 34 | 1,350,310 | 77,658,938 |
| 22 | cit-Cora-KONECT | 23,166 | 89,157 | 7.6972 | 1 | 5 | 16 | 377 | 0.2660 | 0.1268 | 20 | 78,791 | 1,786,074 |
| 23 | cit-HepPh-d-SNAP | 34,401 | 420,784 | 24.4635 | 3 | 15 | 54 | 846 | 0.2856 | 0.1615 | 14 | 1,276,859 | 22,468,237 |
| 24 | cit-HepTh-d-SNAP | 27,400 | 352,021 | 25.6950 | 3 | 15 | 57 | 2,468 | 0.3139 | 0.1299 | 15 | 1,478,698 | 32,665,296 |
| 25 | cit-Patents-d-SNAP | 3,764,117 | 16,511,740 | 8.7732 | 1 | 6 | 19 | 793 | 0.0758 | 0.0703 | 26 | 7,514,922 | 313,229,094 |
| 26 | comm-CALL-ND | 4,295,638 | 7,893,769 | 3.6753 | 1 | 3 | 7 | 110 | 0.2179 | 0.1985 | 45 | 2,253,963 | 31,804,482 |
| 27 | comm-EmailEnron-SNAP | 33,696 | 180,811 | 10.7319 | 1 | 3 | 19 | 1,383 | 0.5092 | 0.0903 | 13 | 725,311 | 23,384,268 |
| 28 | comm-EmailEuAll-SNAP | 32,430 | 54,397 | 3.3547 | 1 | 1 | 3 | 623 | 0.1127 | 0.0273 | 9 | 48,992 | 5,341,634 |
| 29 | comm-LinuxKernel-KONECT | 10,857 | 76,317 | 14.0586 | 1 | 2 | 24 | 1,927 | 0.3486 | 0.1185 | 13 | 698,240 | 16,977,912 |
| 30 | comm-Mobile-ND | 5,324,963 | 10,410,903 | 3.9102 | 1 | 3 | 8 | 22,224 | 0.1811 | 0.0104 | 36 | 2,895,897 | 835,604,293 |
| 31 | comm-SMS-ND | 2,369,078 | 3,330,086 | 2.8113 | 1 | 2 | 6 | 22,224 | 0.0669 | 0.0013 | 42 | 326,282 | 770,920,401 |
| 32 | comm-WikiTalk-SNAP | 92,117 | 360,767 | 7.8328 | 1 | 1 | 11 | 1,220 | 0.0589 | 0.0483 | 11 | 836,467 | 51,083,880 |
| 33 | loc-Brightkite-SNAP | 56,739 | 212,945 | 7.5061 | 1 | 2 | 16 | 1,134 | 0.1734 | 0.1193 | 18 | 494,408 | 11,938,424 |
| 34 | loc-Foursquare-ASU | 639,014 | 3,214,986 | 10.0623 | 1 | 1 | 19 | 106,218 | 0.1080 | 0.0016 | 4 | 21,651,003 | 39,400,700,856 |
| 35 | loc-FourSquare-NetRep | 639,014 | 3,214,986 | 10.0623 | 1 | 1 | 19 | 106,218 | 0.1080 | 0.0016 | 4 | 21,651,003 | 39,400,700,856 |
| 36 | loc-Gowalla-SNAP | 196,591 | 950,327 | 9.6681 | 1 | 3 | 20 | 14,730 | 0.2367 | 0.0239 | 16 | 2,273,138 | 283,580,626 |
| 37 | soc-Academia-NetRep | 137,969 | 369,692 | 5.3591 | 1 | 2 | 12 | 702 | 0.1421 | 0.0866 | 21 | 220,641 | 7,995,452 |
| 38 | soc-Advogato-KONECT | 2,716 | 7,773 | 5.7239 | 1 | 3 | 14 | 138 | 0.2233 | 0.1325 | 13 | 5,383 | 116,510 |
| 39 | soc-BuzzNet-ASU | 101,163 | 2,763,066 | 54.6260 | 2 | 14 | 98 | 64,289 | 0.2321 | 0.0108 | 5 | 30,919,848 | 8,542,533,935 |
| 40 | soc-Catster-NetRep | 148,826 | 5,447,464 | 73.2058 | 3 | 22 | 84 | 80,634 | 0.3877 | 0.0111 | 10 | 185,462,078 | 50,059,386,906 |
| 41 | soc-Delicious-ASU | 536,108 | 1,365,961 | 5.0958 | 1 | 1 | 10 | 3,216 | 0.0322 | 0.0106 | 14 | 487,972 | 137,770,815 |
| 42 | soc-Digg-ASU | 770,799 | 5,907,132 | 15.3273 | 1 | 2 | 16 | 17,643 | 0.0881 | 0.0482 | 18 | 62,710,792 | 3,842,962,151 |
| 43 | soc-Dogster-NetRep | 426,485 | 8,543,321 | 40.0639 | 2 | 12 | 58 | 46,503 | 0.1710 | 0.0144 | 11 | 83,499,945 | 17,303,939,974 |
| 44 | soc-Douban-ASU | 154,908 | 327,162 | 4.2240 | 1 | 1 | 5 | 287 | 0.0161 | 0.0104 | 9 | 40,612 | 11,623,280 |
| 45 | soc-Epinions1-d-SNAP | 75,877 | 405,739 | 10.6947 | 1 | 2 | 18 | 3,044 | 0.1378 | 0.0687 | 15 | 1,624,481 | 69,327,677 |
| 46 | soc-Facebook-MPI | 63,392 | 816,886 | 25.7725 | 1 | 11 | 69 | 1,098 | 0.2218 | 0.1639 | 15 | 3,501,534 | 60,606,675 |
| 47 | soc-Flickr-AMiner | 214,424 | 9,114,421 | 85.0131 | 2 | 23 | 210 | 10,486 | 0.1464 | 0.0832 | 10 | 132,139,697 | 4,630,544,599 |
| 48 | soc-Flickr-ASU | 80,513 | 5,899,882 | 146.5570 | 5 | 46 | 364 | 5,706 | 0.1652 | 0.2142 | 6 | 271,601,326 | 3,531,448,904 |
| 49 | soc-Flickr-MPI | 1,624,992 | 15,476,835 | 19.0485 | 1 | 2 | 19 | 27,236 | 0.1892 | 0.1212 | 24 | 548,646,525 | 13,028,541,364 |
| 50 | soc-Flixter-ASU | 2,523,386 | 7,918,801 | 6.2763 | 1 | 1 | 7 | 1,474 | 0.0834 | 0.0138 | 8 | 7,897,122 | 1,711,880,027 |
| 51 | soc-Friendster-ASU | 5,689,498 | 14,067,887 | 4.9452 | 1 | 1 | 5 | 4,423 | 0.0502 | 0.0048 | 9 | 8,722,131 | 5,484,816,732 |
| 52 | **soc-Friendster-SNAP** | 65,608,366 | 1,806,067,135 | 55.0560 | 1 | 9 | 148 | 5,214 | 0.1623 | 0.0176 | 37 | 4,173,724,142 | 708,133,792,538 |
| 53 | soc-GooglePlus-NetRep | 78,723 | 319,999 | 8.1297 | 1 | 2 | 19 | 538 | 0.1982 | 0.2934 | 59 | 1,386,340 | 12,787,287 |
| 54 | soc-Hamsterster-KONECT | 2,000 | 16,098 | 16.0980 | 2 | 9 | 36 | 273 | 0.5401 | 0.2709 | 10 | 52,665 | 530,614 |
| 55 | soc-Hyves-ASU | 1,402,673 | 2,777,419 | 3.9602 | 1 | 1 | 7 | 31,883 | 0.0448 | 0.0016 | 10 | 752,401 | 1,444,870,827 |
| 56 | soc-LastFM-AMiner | 135,876 | 1,685,158 | 24.8044 | 1 | 9 | 54 | 3,137 | 0.1983 | 0.0946 | 12 | 9,097,399 | 279,291,397 |
| 57 | soc-LastFM-ASU | 1,191,805 | 4,519,330 | 7.5840 | 1 | 2 | 11 | 5,150 | 0.0727 | 0.0111 | 10 | 3,946,207 | 898,270,114 |
| 58 | soc-Libimseti-KONECT | 34,339 | 124,722 | 7.2642 | 1 | 2 | 16 | 613 | 0.0224 | 0.0265 | 15 | 54,375 | 6,103,992 |
| 59 | soc-LinkedIn-AMiner | 6,725,712 | 19,360,071 | 5.7570 | 2 | 4 | 11 | 869 | 0.3700 | 0.2863 | 32 | 12,862,009 | 121,917,817 |
| 60 | soc-LiveJournal1-d-SNAP | 4,843,953 | 42,845,684 | 17.6904 | 1 | 5 | 42 | 20,333 | 0.2743 | 0.1280 | 20 | 285,688,896 | 6,412,296,576 |
| 61 | soc-LiveJournal-AMiner | 3,017,282 | 85,654,975 | 56.7762 | 4 | 17 | 114 | 910,088 | 0.1196 | 0.0017 | 8 | 507,338,233 | 919,635,317,380 |
| 62 | soc-LiveJournal-ASU | 2,238,731 | 12,816,184 | 11.4495 | 1 | 2 | 21 | 5,873 | 0.1270 | 0.0230 | 8 | 28,204,049 | 3,658,174,479 |
| 63 | soc-LiveJournal-MPI | 5,189,809 | 48,688,097 | 18.7630 | 1 | 6 | 45 | 15,017 | 0.2749 | 0.1352 | 23 | 310,784,163 | 6,586,074,658 |
| 64 | soc-LiveJournal-SNAP | 3,997,962 | 34,681,189 | 17.3494 | 1 | 6 | 42 | 14,815 | 0.2843 | 0.1368 | 21 | 177,820,130 | 3,722,307,805 |
| 65 | soc-LiveMocha-ASU | 104,103 | 2,193,083 | 42.1329 | 2 | 13 | 91 | 2,980 | 0.0454 | 0.0142 | 6 | 3,361,651 | 706,231,197 |
| 66 | soc-MySpace-AMiner | 853,360 | 5,635,236 | 13.2072 | 1 | 6 | 26 | 25,105 | 0.0433 | 0.0022 | 14 | 1,256,533 | 1,686,861,075 |
| 67 | soc-Orkut-NetRep | 2,997,166 | 106,349,209 | 70.9665 | 7 | 42 | 152 | 27,466 | 0.1700 | 0.0439 | 9 | 524,643,952 | 35,294,034,217 |
| 68 | soc-Orkut-SNAP | 3,072,441 | 117,185,083 | 76.2814 | 8 | 45 | 162 | 33,313 | 0.1666 | 0.0424 | 9 | 627,584,181 | 43,742,714,028 |
| 69 | soc-Pokec-d-SNAP | 1,632,803 | 22,301,964 | 27.3174 | 1 | 13 | 70 | 14,854 | 0.1094 | 0.0483 | 14 | 32,557,458 | 1,988,401,184 |
| 70 | soc-Prosper-d-KONECT | 89,171 | 3,329,970 | 74.6873 | 3 | 34 | 182 | 6,515 | 0.0049 | 0.0031 | 8 | 1,158,669 | 1,108,949,447 |
| 71 | soc-Slashdot0811-d-SNAP | 77,360 | 469,180 | 12.1298 | 1 | 2 | 25 | 2,539 | 0.0555 | 0.0246 | 12 | 551,724 | 66,861,129 |
| 72 | soc-Slashdot0902-d-SNAP | 82,168 | 504,230 | 12.2731 | 1 | 2 | 25 | 2,552 | 0.0603 | 0.0245 | 13 | 602,592 | 73,175,813 |
| 73 | soc-WikiVote-SNAP | 7,066 | 100,736 | 28.5129 | 1 | 4 | 82 | 1,065 | 0.1419 | 0.1369 | 7 | 608,389 | 12,720,410 |
| 74 | **soc-YouTube-MPI** | 1,134,890 | 2,987,624 | 5.2650 | 1 | 1 | 8 | 28,754 | 0.0808 | 0.0062 | 24 | 3,056,386 | 1,465,313,402 |
| 75 | web-BaiduBaike-KONECT | 2,107,689 | 16,996,139 | 16.1277 | 1 | 4 | 29 | 97,848 | 0.1171 | 0.0025 | 20 | 25,206,270 | 30,809,207,121 |
| 76 | web-BerkStan-SNAP | 654,782 | 6,581,871 | 20.1040 | 2 | 8 | 36 | 84,230 | 0.6066 | 0.0069 | 208 | 64,520,617 | 27,786,200,608 |
| 77 | web-Google-SNAP | 855,802 | 4,291,352 | 10.0288 | 1 | 5 | 20 | 6,332 | 0.5190 | 0.0552 | 24 | 13,356,298 | 486,129,576 |
| 78 | web-Hudong-KONECT | 1,962,418 | 14,419,760 | 14.6959 | 1 | 5 | 23 | 61,440 | 0.0783 | 0.0035 | 16 | 21,611,635 | 18,660,819,412 |
| 79 | web-Stanford-SNAP | 255,265 | 1,941,926 | 15.2150 | 2 | 6 | 29 | 38,625 | 0.6189 | 0.0086 | 164 | 11,277,977 | 3,907,779,392 |
| 80 | web-WWW-ND | 325,729 | 1,090,108 | 6.6933 | 1 | 2 | 12 | 10,721 | 0.2346 | 0.0931 | 46 | 8,910,005 | 278,151,159 |